\documentclass[runningheads]{llncs}
\usepackage[T1]{fontenc}
\usepackage{graphicx}
\usepackage{booktabs}
\usepackage{algorithm}
\usepackage{algpseudocode}
\usepackage{amsmath}
\makeatletter
\let\prismoldthebibliography\thebibliography
\renewcommand{\thebibliography}[1]{%
  \prismoldthebibliography{#1}%
  \small
  \setlength{\itemsep}{0pt}%
  \setlength{\parskip}{0pt}%
}
\makeatother
\begin{document}
\title{An Empirical Study of Multi-Agent Collaboration for Automated Research}
%
%


\author{Yang Shen\inst{1,2}\orcidID{0000-0002-7917-8201} \and
Zhenyi Yi\inst{2}\orcidID{0009-0005-1536-3626} \and
Ziyi Zhao\inst{1}\orcidID{0000-0003-3537-8065} \and
Lijun Sun\inst{3}\orcidID{0000-0002-9830-1527} \and
Dongyang Li\inst{4}\orcidID{0000-0001-6197-5950} \and
Chin-Teng Lin\inst{1}\orcidID{0000-0001-8371-8197} \and
Yuhui Shi\inst{2}\orcidID{0000-0002-8840-723X}}
\authorrunning{Yang Shen et al.}
%
\institute{University of Technology Sydney, Ultimo NSW 2007, Australia 
\and
Southern University of Science and Technology, Shenzhen 518055, China
\email{shiyh@sustech.edu.cn}
\and
Shenzhen Technology University, Shenzhen 518118, China \\
\and
Tongji University, Shanghai 201804, China \\
}
\maketitle              
\begin{abstract}
As AI agents evolve, the community is rapidly shifting from single Large Language Models (LLMs) to Multi-Agent Systems (MAS) to overcome cognitive bottlenecks in automated research. However, the optimal multi-agent coordination framework for these autonomous agents remains largely unexplored. In this paper, we present a systematic empirical study investigating the comparative efficacy of distinct multi-agent structures for automated machine learning optimization. Utilizing a rigorously controlled, execution-based testbed equipped with Git worktree isolation and explicit global memory, we benchmark a single-agent baseline against two multi-agent paradigms: a subagent architecture (parallel exploration with post-hoc consolidation) and an agent team architecture (experts with pre-execution handoffs). By evaluating these systems under strictly fixed computational time budgets, our findings reveal a fundamental trade-off between operational stability and theoretical deliberation. The subagent mode functions as a highly resilient, high-throughput search engine optimal for broad, shallow optimizations under strict time constraints. Conversely, the agent team topology exhibits higher operational fragility due to multi-author code generation but achieves the deep theoretical alignment necessary for complex architectural refactoring given extended compute budgets. These empirical insights provide actionable guidelines for designing future autoresearch systems, advocating for dynamically routed architectures that adapt their collaborative structures to real-time task complexity.

\end{abstract}
\section{Introduction}

In the pursuit of Artificial General Intelligence (AGI), endowing large language model (LLM) based agents with the autonomous capability to conduct scientific research and discover novel knowledge remains a pivotal milestone. This vision has catalyzed the emergence of {\it autoresearch}. Recently, automated research frameworks have demonstrated the feasibility of a fully autonomous machine learning research loop. A paradigmatic example is Karpathy's `autoresearch` project~\cite{karpathy2026autoresearch}, which establishes a loop in which an LLM proposes a configuration change, executes training, and accepts or rejects the change based on validation loss. When repeated, this propose-execute-evaluate loop constitutes a basic form of automated research~\cite{qu2026bilevel}.

However, as research tasks become more complex, monolithic single-agent architectures expose critical structural deficiencies. Single LLMs struggle to maintain long-horizon context~\cite{yao2022react,liu2024lost}, mitigate hallucinations during deep code refactoring, and escape deterministic search paths. While human-engineered extensions have attempted to address these limitations, such as introducing multi-batch parallel search~\cite{liu2026autoresearchclaw} or persistent experience memory~\cite{lyu2026evoscientist}, the fundamental constraints of a single-agent bottleneck persist. To overcome these limitations, the community is rapidly shifting toward Multi-Agent Systems (MAS). In the specific context of automated machine learning research, which is highly dynamic and empirically driven, a critical open question remains: {\it how should multi-agent collaboration topologies be organized to maximize research efficiency?~\cite{liu2023agentbench}}

To address this gap, we present a systematic empirical study of multi-agent coordination frameworks (Fig.~\ref{fig1}). We compare three architectures under fixed training budgets: a single-agent baseline that follows the original `autoresearch` loop, a parallel multi-agent system with post-training coordination, and a pre-training collaboration architecture based on fixed-role expert agents.

Our primary contributions are as follows: 1) A rigorous evaluation testbed for automated research: we build a reproducible, execution-based sandbox for budget-constrained machine learning optimization. With Git Worktree isolation, a deterministic "Search/Replace" patch contract, and explicit global memory, the testbed reduces state contamination and context degradation, enabling fair comparisons under fixed compute budgets. 2) A systematic comparison of multi-agent topologies: we compare distinct organizational structures for neural network optimization and directly study how communication timing and role distribution affect performance. 3) Empirical design guidelines: we quantify how different topologies shape search trajectories, crash rates, proposal validity, and $\texttt{val\_bpb}$, yielding practical guidance for more stable and efficient autoresearch.

\section{Related Work}



A major milestone in this domain is the AI Scientist~\cite{lu2024ai} by Sakana AI, which proposes an end-to-end LLM-driven scientific discovery process. The AI Scientist extends the autonomous loop to encompass the entire scientific lifecycle, including idea generation, literature review, experiment execution, manuscript writing, and automated peer review. While The AI Scientist demonstrates the remarkable potential of LLMs to autonomously produce full academic papers, its underlying execution engine relies heavily on a linear, single-agent pipeline. Fully Automated Research System (FARS)\footnote{\url{https://analemma.ai/fars/}} is an end-to-end AI research system designed to autonomously complete the research workflow without human intervention. FARS is implemented as a multi-agent research system, including four specialized agents: ideation, planning, experiment, and writing. FARS ran for 417 hours, consuming 21.6 billion tokens to generate 166 papers, at a total cost of \$186,000 USD. The AI Scientist and FARS are not affordable for the vast majority of researchers. Generic agent frameworks and collaboration paradigms have been well studied.~\cite{yao2022react,wei2022chain,park2023generative}

Autoresearch~\cite{karpathy2026autoresearch} deploys a single AI agent running research on single-GPU training automatically, which is super-lightweight. Subsequent works have expanded on this base. AutoResearch Claw~\cite{liu2026autoresearchclaw} extends this framework with multi-batch parallelism by evaluating several candidate configurations simultaneously, while EvoScientist~\cite{lyu2026evoscientist} introduces persistent experience memory to enable cross-run learning. In addition, ARIS~\cite{yang2026aris} introduces cross-model review mechanisms to mitigate the evaluation blind spots inherent in single-model systems. Concurrently, ArgusBot~\cite{fan2026argusbot} addresses long-horizon execution bottlenecks by implementing a continuous supervision architecture that distributes tasks across specialized planning, execution, and review roles. Recent work Bilevel Autoresearch~\cite{qu2026bilevel} takes a meta-optimization approach, utilizing an outer loop to generate and inject new search mechanisms as Python code at runtime to break deterministic search patterns. While these works focus on algorithm generation, batching, or meta-optimization, our work highlights multi-agent autoresearch. 

Previous work~\cite{yang2020swarm} has mentioned the challenge of swarm intelligence-based AutoML, while current agents are good at working independently on specific tasks, but are often overwhelmed when dealing with complex engineering projects that require teamwork. Besides, the limitation of single-agent sequential reasoning/execution is heavily dependent on a single prompt, lacking the ability to brainstorm~\cite{shen2020bso}. Brainstorming can enable a group of agents to interact and suggest ideas spontaneously in response to a prompt. To address this, many foundation model vendors are now supporting agent swarms and emphasizing the importance of agents' teamwork. For example, Kimi K2.5 introduces the agent swarm to address the limitations of single-agent sequential reasoning/execution~\cite{team2026kimi}, and Kimi K2.6 can even run continuously for 12 hours with 300+ agents in parallel from a single prompt. Claude Code now also supports agent teams for complex work.

All existing autoresearch works either focus on single-agent automation or on role-based multi-agent research systems and lack inter-agent communication, which plays a crucial role~\cite{sun2025multi} in multi-agent systems. In this paper, we build on the autoresearch loop for neural network hyperparameter search and extend it to a multi-agent version under fixed computational budgets.

\section{Methods}

To systematically investigate how multi-agent organizational structures impact autoresearch efficiency, we formulate the automated research process as a constrained search optimization problem. We construct a rigorous evaluation testbed designed to isolate the effects of collaboration topologies while neutralizing confounding variables such as state contamination and catastrophic forgetting.

\subsection{Task Formulation}

We define the automated machine learning research task as the iterative optimization of a target codebase (specifically, the \texttt{train.py} script). The objective function is to minimize the validation bits per byte (\texttt{val\_bpb} denoted as $\mathcal{L}_{val}$). Let $\mathcal{P}_0$ be the initial state of the codebase. In each iteration $t$, the agent system proposes a patch $\Delta \mathcal{P}_t$. The modification is accepted if and only if $\mathcal{L}_{val}(\mathcal{P}_{t-1} + \Delta \mathcal{P}_t) < \mathcal{L}_{val}(\mathcal{P}_{t-1})$. Crucially, to benchmark candidate-level evaluation under controlled training cost, each candidate training job is assigned a fixed training budget $\mathcal{T}_{train}$, evaluated at $\mathcal{T}_{train}=300s$ and $\mathcal{T}_{train}=600s$ in our experiments. All methods are then compared over the same number of research rounds, with additional analysis by proposal counts, failure counts, and cumulative training time.

\subsection{Evaluation Testbed and Experimental Controls}

Environmental contamination~\cite{jimenez2023swe} and context degradation~\cite{liu2024lost,lu2024ai} are the two most common issues when evaluating autonomous code generation agents. To ensure fair comparisons across different multi-agent topologies, we set the following experimental rules.

{\bf Git Worktree Isolation} To support concurrent exploration without cross-contamination, the testbed dynamically allocates completely isolated Git worktrees for each agent or shared candidate. All divergent exploration branches originate from an identical baseline commit in a given iteration.

{\bf Structured Patch Contract} To prevent agents from destructively overwriting entire source files or losing executable structures, interactions with the codebase are restricted to a strict Search/Replace contract. Agents must output a structured proposal containing a theoretical \texttt{motivation}, an \texttt{idea\_summary}, and precise \texttt{search\_block/replace\_block} pairs.

{\bf Preflight Validation} Before consuming the actual training time budget, all generated patches undergo a lightweight preflight compilation check to intercept glaring syntax errors and prevent trivial runtime crashes.

{\bf Explicit Global Memory Base} Catastrophic forgetting is very common in long-horizon agentic tasks. To isolate the collaborative topology as the primary variable, we equip the testbed with a baseline explicit memory mechanism applied uniformly across all evaluated architectures. A shared document (\texttt{program\_exp.md}) records experiences incrementally, such as mechanisms that previously yielded improvements or induced crashes, preventing agents from hallucinating historical facts or falling into cyclic failure loops.

\subsection{Evaluated Multi-Agent Coordination Frameworks}

\begin{figure}[htbp]
\centering
\includegraphics[width=0.97\textwidth]{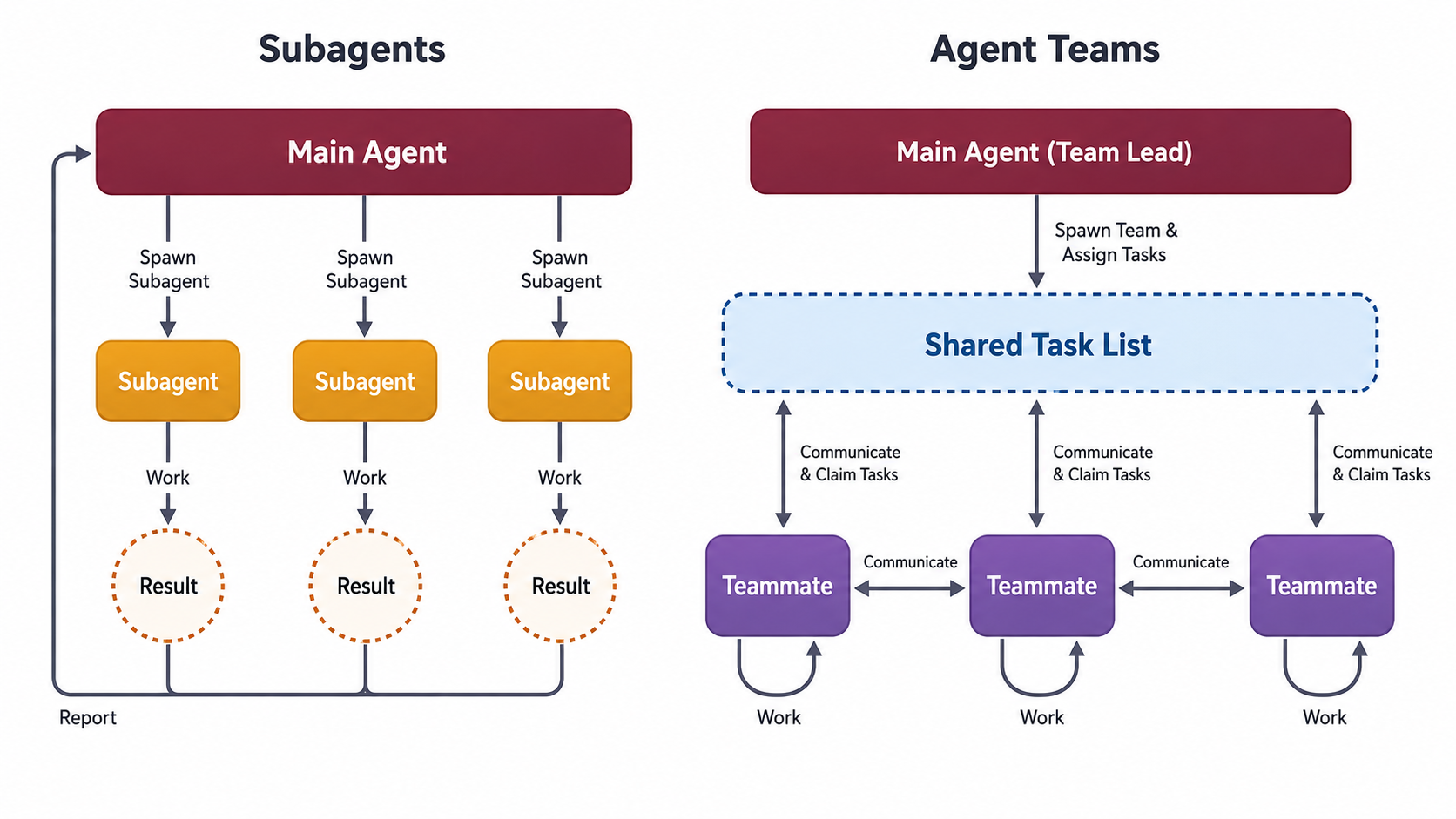}
\caption{Multi-Agent Coordination Frameworks} 
\label{fig1}
\end{figure}

In this paper, we investigate two of the most popular multi-agent architectures, i.e., subagents and agent teams, as shown in Fig.~\ref{fig1}.

The {\bf Subagents} architecture follows a hierarchical design in which a central agent (the orchestrator) decomposes a complex task into subtasks and delegates them to specialized subagents. Each subagent operates independently with its own context and prompt, focusing on a specific function such as code generation, testing, or debugging. The orchestrator maintains global control by coordinating execution order, aggregating intermediate results, and ensuring consistency across outputs. This design provides strong controllability and modularity, making it well-suited for structured workflows where task decomposition is clear and predictable. However, because subagents do not directly communicate with one another, all interactions must pass through the orchestrator, which can limit flexibility and introduce coordination overhead.

This topology evaluates the efficacy of distributing cognitive load through parallel search, and then a centralized coordinator agent merges the modifications by multiple subagents. The procedure of each round is described as follows. First, multiple worker agents independently read the context and generate distinct proposals in isolated worktrees, and execute candidate training jobs sequentially on the same GPU. Second, if more than one candidate successfully improves upon the baseline metric within the same iteration, a coordinator agent is triggered. The Coordinator attempts to merge the distinct, high-potential patches into a single unified script. Finally, the merged candidate undergoes an independent evaluation. The merged solution will only be accepted if its performance strictly surpasses that of the best agent in that round, ensuring that the merging process does not degrade isolated gains.

\begin{algorithm}[!htb]
\caption{Subagent Procedure}
\label{alg:sub}
\scriptsize
\begin{algorithmic}[1]
\Require Initial codebase $P_0$, validation loss function $\mathcal{L}_{val}$, number of rounds $R_{max}$, candidate training budget $\mathcal{T}_{train}$
\Require Global explicit memory $\mathcal{M}_{exp}$, number of workers $K$
\Ensure Optimized codebase $P_{main}$

\State $P_{main} \gets P_0$
\State $L_{best} \gets \mathcal{L}_{val}(P_{main})$
\For{$t = 1$ \textbf{to} $R_{max}$}
    \State $Candidates \gets \emptyset$

    \For{$k = 1$ \textbf{to} $K$}
        \State $W_k \gets \text{CreateIsolatedWorktree}(P_{main})$
        \State $Patch_k, Intent_k \gets \text{LLM\_Worker}(W_k, \mathcal{M}_{exp})$
        \If{$\text{PreflightCheck}(W_k \oplus Patch_k)$ is SUCCESS}
            \State $L_k \gets \text{Train}(W_k \oplus Patch_k, \mathcal{T}_{train})$
            \If{$L_k < L_{best}$}
                \State $Candidates\text{.Append}(\{patch: Patch_k, loss: L_k\})$
            \Else
                \State $\text{UpdateMemory}(\mathcal{M}_{exp}, Intent_k, \text{"Failed"})$
            \EndIf
        \EndIf
    \EndFor

    \If{$|Candidates| > 1$}
        \State $BestIndiv \gets \arg\min_{c \in Candidates} c.loss$
        \State $W_{merged} \gets \text{CreateIsolatedWorktree}(P_{main})$
        \State $Patch_{merged} \gets \text{LLM\_Coordinator}(Candidates)$
        \If{$\text{PreflightCheck}(W_{merged} \oplus Patch_{merged})$ is SUCCESS}
            \State $L_{merged} \gets \text{Train}(W_{merged} \oplus Patch_{merged}, \mathcal{T}_{train})$
            \If{$L_{merged} < BestIndiv.loss$}
                \State $Candidates\text{.Append}(\{patch: Patch_{merged}, loss: L_{merged}\})$
            \EndIf
        \EndIf
    \EndIf

    \If{$|Candidates| > 0$}
        \State $BestCand \gets \arg\min_{c \in Candidates} c.loss$
        \State $P_{main} \gets P_{main} \oplus BestCand.patch$
        \State $L_{best} \gets BestCand.loss$
        \State $\text{UpdateMemory}(\mathcal{M}_{exp}, BestCand.patch, \text{"Success"})$
    \EndIf
\EndFor

\State \Return $P_{main}$
\end{algorithmic}
\end{algorithm}

Algorithm~\ref{alg:sub} formalizes the subagent procedure described above. In each round, multiple workers independently propose patches from isolated worktrees, while candidate training jobs are executed sequentially on the same GPU. When multiple worker candidates improve over the current baseline, the coordinator attempts to merge their patches; the merged candidate is accepted only if it strictly outperforms the best worker candidate in that round.

\begin{algorithm}[!t]
\caption{Agent Teams Procedure}
\label{alg:team}
\scriptsize
\begin{algorithmic}[1]
\Require Initial codebase $P_0$, validation loss $\mathcal{L}_{val}$, number of rounds $R_{max}$, candidate training budget $\mathcal{T}_{train}$
\Require Global memory $\mathcal{M}_{exp}$, meta-memory $\mathcal{M}_{meta}$
\Require Role sequence $\mathcal{R}=[A,O,E,A,O,E]$
\Ensure Optimized codebase $P_{main}$

\State $P_{main} \gets P_0$
\State $L_{best} \gets \mathcal{L}_{val}(P_{main})$
\For{$t = 1$ \textbf{to} $R_{max}$}
    \State $W_{shared} \gets \text{CreateIsolatedWorktree}(P_{main})$
    \State $Ctx \gets \emptyset$

    \For{\textbf{each} $Role \in \mathcal{R}$}
        \State $Patch, Summary \gets \text{LLM\_Expert}(Role, W_{shared}, Ctx)$
        \State $W_{shared} \gets \text{ApplyPatch}(W_{shared}, Patch)$
        \State $Ctx \gets \text{AppendContext}(Ctx, Summary)$
    \EndFor

    \If{$\text{PreflightCheck}(W_{shared})$ is FAILED}
        \State $\text{UpdateMemory}(\mathcal{M}_{meta}, Ctx, \text{"Preflight failure"})$
        \State \textbf{continue}
    \EndIf

    \State $Status, L_{shared}, ErrorLog \gets \text{TrainAndEvaluate}(W_{shared}, \mathcal{T}_{train})$
    \If{$Status$ is CRASH}
        \State $W_{repair}, fixed \gets \text{LLM\_Engineer}(W_{shared}, ErrorLog, Ctx)$
        \If{\textbf{not} $fixed$ \textbf{or} $\text{PreflightCheck}(W_{repair})$ is FAILED}
            \State $\text{UpdateMemory}(\mathcal{M}_{meta}, Ctx, \text{"Unresolvable crash"})$
            \State \textbf{continue}
        \EndIf
        \State $Status, L_{shared}, ErrorLog \gets \text{TrainAndEvaluate}(W_{repair}, \mathcal{T}_{train})$
        \If{$Status$ is CRASH}
            \State $\text{UpdateMemory}(\mathcal{M}_{meta}, Ctx, \text{"Unresolvable crash"})$
            \State \textbf{continue}
        \EndIf
        \State $W_{shared} \gets W_{repair}$
    \EndIf

    \If{$L_{shared} < L_{best}$}
        \State $P_{main} \gets W_{shared}$
        \State $L_{best} \gets L_{shared}$
        \State $\text{UpdateMemory}(\mathcal{M}_{exp}, Ctx, \text{"Success"})$
    \Else
        \State $\text{UpdateMemory}(\mathcal{M}_{exp}, Ctx, \text{"Failed"})$
    \EndIf
\EndFor

\State \Return $P_{main}$
\end{algorithmic}
\end{algorithm}

\vspace{-0.4em}

This topology evaluates pre-training collaboration and fixed-role specialization. The system maintains only a single shared worktree per iteration, focusing computing resources on the best candidate. In our experiments, we predefined three expert agents: an architecture expert, an optimizer \& schedule expert, and an efficiency \& memory expert. It is worth noting that unrestricted, random communication can lead to a dramatic increase in the crash rate; thus, we adopt a more conservative approach to simulate communication behavior among agents. We let the agents operate in a predefined sequential chain across multiple conversational turns ($6$ turns in our experiment). Rather than merging code by a centralized coordinator, expert agents work together to edit the codebase incrementally. Subsequent expert agents receive the current code state alongside the structural \texttt{idea\_summary} and \texttt{motivation} passed down by other expert agents (we call it group chat). This process is similar to multiple engineers maintaining a GitHub repository. The training occurs only after the group chat concludes. To mitigate the fragility of complex, multi-author code, an isolated Engineer agent serves as a fallback mechanism. In the event of a runtime crash, the Engineer performs conservative debugging to restore executability while strictly preserving the experts' intent. The procedure is shown in Alg.~\ref{alg:team}.

\section{Experiments and Results}

Using the standardized testbed, we conduct three distinct experiments to compare the timing and structure of multi-agent collaboration. All experiments optimize the same target file, \texttt{train.py}, adapted from Karpathy's \texttt{autoresearch} repository\footnote{\url{https://github.com/karpathy/autoresearch}}. The task is next-token prediction on the \texttt{karpathy/climbmix-400b-shuffle} dataset, using fixed HuggingFace parquet shards with \texttt{shard\_06542.parquet} as the validation shard. The evaluation metric is validation bits per byte, \texttt{val\_bpb}, computed by summing the per-token cross entropy over the validation loader, normalizing by the byte length of target tokens, and dividing by $\log 2$.

The optimized \texttt{train.py} is a single-file GPT-style autoregressive Transformer pretraining script. The baseline model has $8$ Transformer blocks, hidden dimension $512$, $4$ attention heads, maximum sequence length $2048$, rotary positional embeddings, RMS normalization, local/global attention windows following the \texttt{SSLL} pattern, and MLP blocks with a $4\times$ expansion ratio. The agents optimize the experiment by proposing source-level patches to \texttt{train.py}, including changes to model architecture, attention/window patterns, MLP width, value embeddings, rotary embedding length, learning-rate schedules, optimizer settings, and other implementation choices that may improve \texttt{val\_bpb}.

Because the original autoresearch setting targets an NVIDIA H100 GPU, we adapt the benchmark to a single NVIDIA RTX 3090 with 24GB memory. The main changes include reducing \texttt{EVAL\_TOKENS} from $40 \times 524288$ to $10 \times 524288$, reducing \texttt{TOTAL\_BATCH\_SIZE} from $2^{19}$ to $2^{17}$ in the 3090 adaptation, and using a medium-sized benchmark configuration with \texttt{MAX\_SEQ\_LEN}=2048, \texttt{DEPTH}=8, \texttt{model\_dim}=512, four attention heads, and \texttt{DEVICE\_BATCH\_SIZE}=32. The benchmark uses fixed random seeds, \texttt{torch.manual\_seed(42)} and \texttt{torch.cuda.manual}\allowbreak \texttt{\_seed(42)}, mixed-precision bfloat16 autocast, and \texttt{torch.compile}. For more details on the experimental configuration, please refer to our code repository\footnote{Code is available at \url{https://github.com/yshenfab/MAAR}}.

We evaluate two candidate-level training budgets, $T=300s$ and $T=600s$, applied to each training job rather than to an entire round or run. To ensure a fair comparison, all three systems share the same benchmark configuration, target file, validation metric, preflight rules, and single-GPU environment within each budget setting. LLM inference, preflight checking, and engineer debugging are excluded from the budget. Because the subagent system can generate up to three worker proposals and one coordinator proposal per round, we report not only rounds but also proposal counts, failure counts, and cumulative training time.

In our experiments, we use \texttt{glm-4.7} and \texttt{glm-4.6v}~\cite{hong2025glm} as agent backbones. The single-agent baseline uses one \texttt{glm-4.6v} worker that observes the main branch codebase and global memory, sequentially generates one proposal, executes training, and retains a modification only if it improves \texttt{val\_bpb}. The subagent mode uses three \texttt{glm-4.6v} workers and a \texttt{glm-4.7} coordinator. The agent teams mode uses three specialist roles, architecture, optimizer/schedule, and efficiency/memory, executed in a fixed six-turn relay, with a \texttt{glm-4.7} engineer triggered only when the final shared candidate crashes during training.

\begin{figure}[t]
\centering
\includegraphics[width=0.97\textwidth]{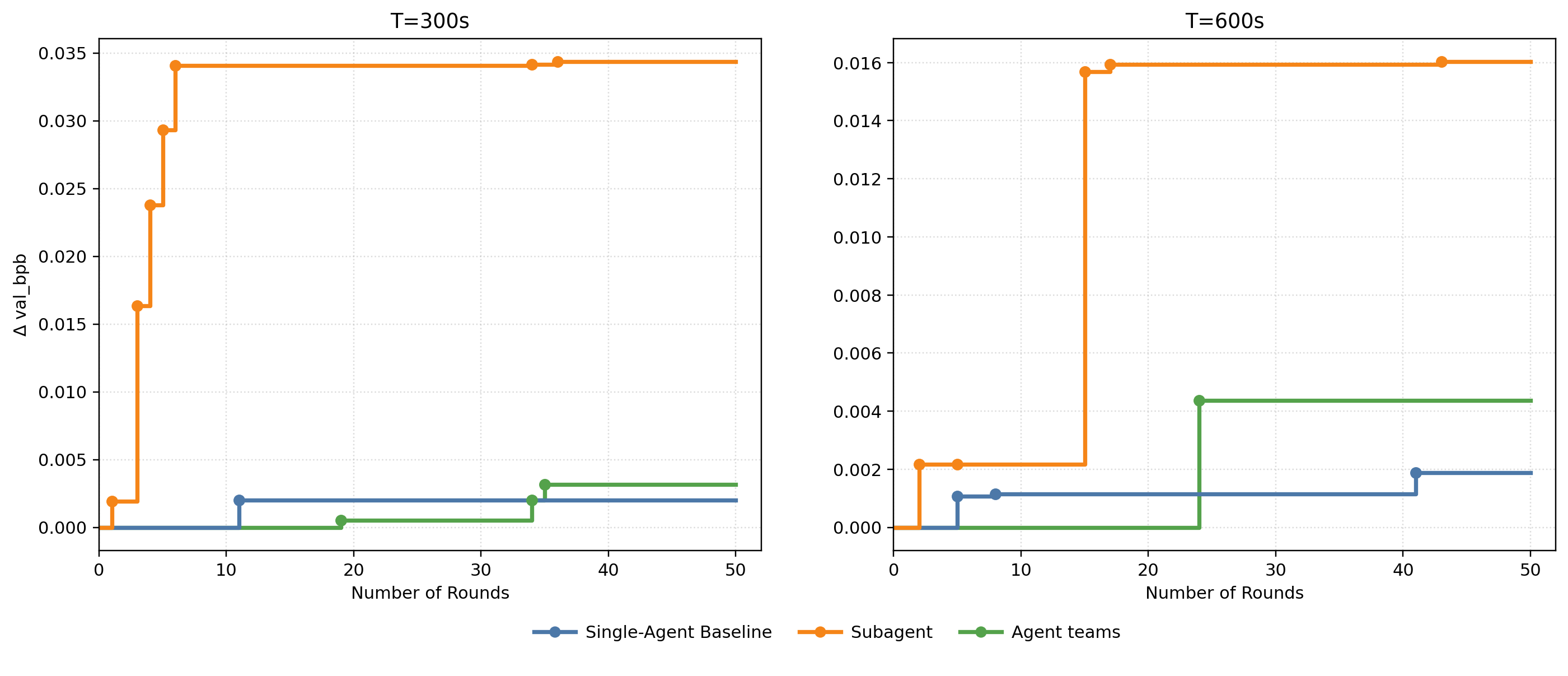}
\caption{Autoresearch Progress}
\label{fig:ar_progress}
\end{figure}

To empirically evaluate the efficacy of the proposed multi-agent topologies, we tracked their ability to improve the target model's performance over successive research rounds. The primary evaluation metric is the absolute reduction in the validation bits per byte, denoted as $\Delta \texttt{val\_bpb}$, where a higher positive value indicates a more successful optimization of the underlying architecture and hyperparameters. The results across two distinct computational budget constraints $\mathcal{T}_{max} = 300s$ (left) and $\mathcal{T}_{max} = 600s$ (right) are shown in Fig.~\ref{fig:ar_progress}. The single-agent baseline shows a moderate initial improvement but then plateaus quickly. Constrained by its linear trial-and-error nature, it struggles to escape local optima once the obvious hyperparameter tweaks are exhausted. Subagent mode demonstrates a distinct early-stage advantage. By parallelizing proposal generation and evaluating multiple candidate directions within each round, it rapidly accumulates localized improvements on $\Delta \texttt{val\_bpb}$. The centralized coordinator successfully merges these independent gains. Conversely, agent teams show a slower initial improvement. Because this topology invests significant context window and API calls into pre-execution deliberation among experts~\cite{yao2023tree,yao2022react}, fewer actual training runs are executed within the short time limit.

In Fig.~\ref{fig:ar_progress} ($T=300s$), the subagent mode has seven effective improvements over $50$ rounds, while the agent teams have only three improvements. However, the subagent modifications exhibited a distinct lack of diversity. The subagents predominantly engaged in greedy local optimization, frequently falling into a repetitive cycle of single-dimensional hyperparameter squeezing. For instance, the subagent system became persistently fixated on iteratively reducing the MLP expansion ratio (e.g., sequentially scaling from $4\times$ down to $0.75\times$ across multiple rounds) rather than exploring other architectural dimensions. In contrast, although the agent teams produced fewer successful improvements, the modifications demonstrated significantly higher diversity and structural complexity. Rather than getting stuck in a single-dimensional parameter trap, the agent teams successfully formulated and executed coupled architectural changes within a single iteration. For example, under the $T=300s$ budget, the team proposed a unified patch that simultaneously adjusted the window attention pattern (from SSLL to SLSL), modified the learning rate warmdown schedule ($0.50 \rightarrow 0.30$), and altered the value embedding vocabulary size. Similarly, for $T=600s$, they executed combined modifications across MLP widths, RoPE lengths, and scheduling. This qualitative evidence indicates that the pre-execution deliberation inherent in the agent team topology fosters a more holistic exploration of the architectural design space, empowering the system to discover complex, multi-faceted improvements that parallel subagents fail to conceptualize.




\begin{figure}[t]
\centering
\includegraphics[width=0.97\textwidth]{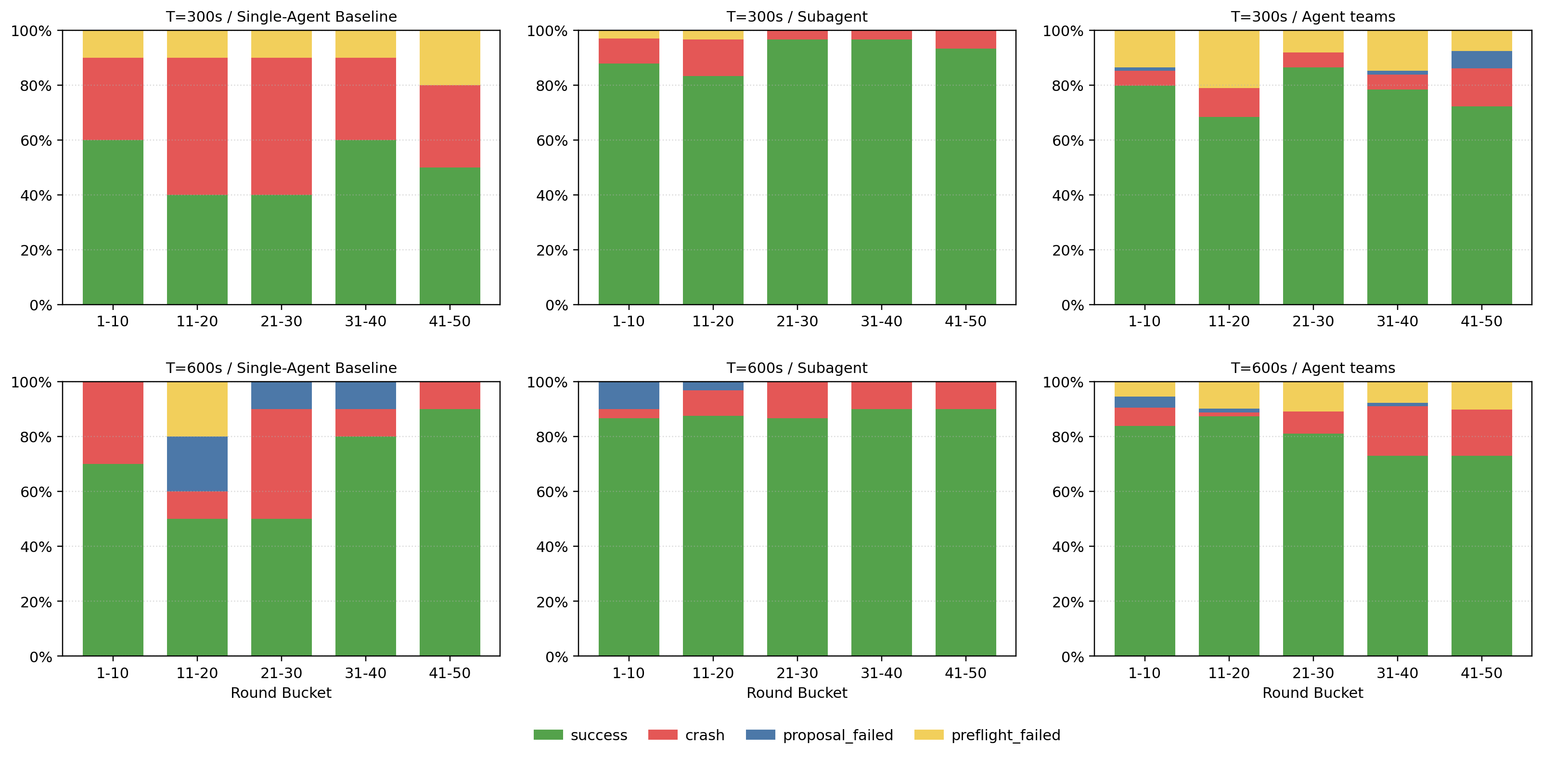}
\caption{Ratio of each phase}
\label{fig:ratio}
\end{figure}

While $\Delta \texttt{val\_bpb}$ measures optimization quality, process stability is equally important. Fig.~\ref{fig:ratio} compares the success profiles of the single-agent, subagent, and agent-teams systems. We track each proposal through four states: Proposal Failure, for patches that violate the Search/Replace format; Preflight Failure, for rule-compliant patches rejected by static checks; Training Crash, for patches that pass preflight but fail at runtime; and Training Success, for patches that complete training and evaluation.

The single-agent baseline, constrained by its linear, single-threaded execution bottleneck, is unable to generate logically sound, complex code without external feedback. Fig.~\ref{fig:ratio} shows that it suffers from significant failures, reflecting the inherent difficulty of zero-shot complex edits, which corresponds to~\cite{jimenez2023swe,liu2023agentbench}.

Subagent mode functions as a stable and efficient empirical search engine. Because workers explore independently in isolated worktrees, individual failures do not halt progress. Fig.~\ref{fig:ratio} confirms this resilience: subagents have the lowest effective preflight-failure and crash rates and generate the largest number of valid proposals within the time budget, making them well suited to broad but structurally simple exploration. Agent teams instead expose a trade-off between deliberation and fragility. As experts sequentially edit the same script through text handoffs (idea plus \texttt{diff}), the risk of incompatible tensor dimensions, missing imports, and logical errors compounds across turns. For our relatively simple AutoML task under a fixed budget, subagents are more stable and efficient, whereas agent teams retain greater upside for complex modifications. Because our RTX 3090 setup is more constrained than the original H100-based setting, these results should be read as a controlled comparison of collaboration topologies rather than a claim about absolute best performance. Detailed comparison is listed in Table~\ref{tab:agent_comparison}.

\begin{table}[htbp]
    \centering
    \small
    \caption{Comparison Between Subagent and Agent Teams}
    \label{tab:agent_comparison}
    \renewcommand{\arraystretch}{1.2}
    \begin{tabular}{@{} l p{0.35\textwidth} p{0.35\textwidth} @{}}
        \toprule
        & \textbf{Subagents} & \textbf{Agent teams} \\
        \midrule
        \textbf{Context}       & sseturn results to the caller & Fully independent \\
        \textbf{Communication} & Subagents report to main & Agents message each other \\
        \textbf{Coordination}  & Main agent as manager & Multi-agents share task list \\
        \textbf{Best for}      & Simple tasks       & Complex works \\
        \textbf{Token cost}    & Lower    & Higher \\
        \bottomrule
    \end{tabular}
\end{table}


\section{Conclusion}

We present a systematic empirical study of multi-agent coordination frameworks for automated machine learning research. Using a controlled execution testbed, we compare a single-agent baseline with subagent and agent-teams architectures under fixed computational budgets. The results reveal a clear trade-off: subagent architectures provide a resilient, high-throughput search mechanism for broad, shallow optimization, whereas agent teams sacrifice execution stability to gain the deeper alignment needed for complex refactoring. These findings suggest that future autoresearch systems should move beyond rigid organizations toward routed multi-agent designs that allocate subagents for broad sweeps and specialist teams for deeper algorithmic changes as task complexity evolves.
%
%
%
\bibliographystyle{splncs04}
\bibliography{ref}
\end{document}